\documentclass[fleqn,usenatbib]{mnras}

\usepackage[T1]{fontenc}
\usepackage{ae,aecompl}
\usepackage{subfigure}
\usepackage{multicol}
\usepackage{hyperref}

\usepackage{graphicx}	
\usepackage{amsmath}	
\usepackage{amssymb}	

\newcommand{\rmms}{\rm{M_{\odot}}}

\newcommand{\rmzs}{\rm{Z_{\odot}}}

\newcommand{\mf}{M_{\rm form}}
\newcommand{\mm}{M_{\rm merg}}
\newcommand{\mr}{M_{\rm ratio}}
\newcommand{\td}{t_{\rm delay}}

\title[Host galaxy evolution]{The host galaxies of double compact objects across cosmic time}

\author[M. Toffano et al.]{Mattia Toffano$^{1,2}$, Michela Mapelli$^{3,4,5,6}$, Nicola Giacobbo $^{3,4,5,6}$, M. Celeste Artale$^5$, 
	\newauthor  Giancarlo Ghirlanda$^2$
\\
$^{1}$Universit\'a degli Studi dell'Insubria, Via Valleggio 11, 22100, Como, Italy\\
$^{2}$INAF-Osservatorio Astronomico di Brera, Via E. Bianchi 46, 23807 Merate, Italy\\
$^{3}$INAF-Osservatorio Astronomico di Padova, Vicolo dell'Osservatorio 5, 35122 Padova, Italy\\
$^{4}$INFN, Milano Bicocca, Piazza della Scienza 3, I-20126 Milano, Italy\\
$^{5}$Institut f\"ur Astro- und Teilchenphysik, Universit\"at Innsbruck, Technikerstrasse 25/8, A-6020, Innsbruck, Austria\\
$^{6}$Dipartimento di Fisica e Astronomia ``G. Galilei'', Universit\'a di Padova, Vicolo dell'Osservatorio 3, I-35122, Italy\\
}

\date{Accepted XXX. Received YYY; in original form ZZZ}

\pubyear{2019}

\begin{document}
\label{firstpage}
\pagerange{\pageref{firstpage}--\pageref{lastpage}}
\maketitle

\begin{abstract}
  We explore the host galaxies of compact-object binaries (black hole--black hole binaries, BHBs; neutron star--black hole binaries, NSBHs; double-neutron stars; DNSs) across cosmic time, by means of population-synthesis simulations combined with the Illustris cosmological simulation. At high redshift ($z\gtrsim{}4$) the host galaxies of BHBs, NSBHs and DNSs are very similar and are predominantly low-mass galaxies (stellar mass $M<10^{11}$ M$_\odot$). If $z\gtrsim{}4$ most compact objects form and merge in the same galaxy, with a short delay time. At low redshift ($z\leq{}2$), the host galaxy populations of DNSs differ significantly from the host galaxies of both BHBs and NSBHs. DNSs merging at low redshift tend to form and merge in the same galaxy, with relatively short delay time. The stellar mass of DNS hosts peaks around $\sim{}10^{10}-10^{11}$ M$_\odot$. In contrast, BHBs and NSBHs merging at low redshift tend to form in rather small galaxies at high redshift and then to merge in larger galaxies with long delay times. This difference between DNSs and black hole binaries is a consequence of their profoundly different metallicity dependence.
\end{abstract}




\section{Introduction}\label{sec:intro}

In September 2015, the two advanced LIGO interferometers \citep{aasi15,harry10} obtained the first direct detection of  gravitational waves (GWs) from a merging black hole--black hole binary (BHB,  \citealt{abbott16b}). This detection was followed by other two BHB events during the first observing run \citep{abbott16a,abbott16c}, while the second observing run brought seven additional BHB mergers \citep{
abbott17a,abbott17b,abbott17f,abbott18a} and one double neutron star (DNS) merger  (GW170817, \citealt{abbott17c}). The advanced Virgo interferometer \citep{acernese15} joined the two advanced LIGO detectors in August 2017, for the last part of O2, allowing for measurements of GW polarization and leading to a dramatic improvement of sky localization \citep{abbott17b,abbott17c,abbott18a,abbott19}. The third observing run of the LIGO--Virgo collaboration (LVC) has just started (April 1st 2019): by the end of O3, we expect several tens of BHBs and few additional DNSs to enrich the current population of GW events.

GW170817 is the only event for which an electromagnetic counterpart was detected, covering almost the whole electromagnetic spectrum, from gamma  to radio wavelengths \citep{abbott17d, abbott17e, abdalla17, alexander17, chornock17, coulter17, cowperthwaite17, ghirlanda19, goldstein17, hallinan17, margutti17, mooley18, nicholl17, pian17, savchenko17, smartt17, soares17,troja17,valenti17}.
 The counterpart has allowed to uniquely identify the host galaxy as NGC~4993 \citep{coulter17}, an early-type massive galaxy at $\sim{}40$~Mpc distance from us \citep{levan17,im17,fong17,blanchard2017}. 

Shedding light on the properties of the host galaxies of black hole (BH)  and neutron star (NS) binaries is a crucial step towards our understanding of the GW Universe (see e.g.  \citealt{cao18, lamberts18, dvorkin16, lamberts16, map17, oshaughnessy17, schneider17, elbert18, mapgiac18,artale19}). 
A good grasp on the properties of the host galaxies  would enable us to test and improve different models of compact-binary formation, which are still affected by a plethora of uncertainties (see e.g. \citealt{tutukov73, flannery75, bethe98, portegies98, portegies00, belczynski02, voss03, podsiadlowski04, tauris06, belczynski07, bogomazov07, dominik12, dominik13, dominik15, mennekens14, demink15, spera15, tauris15, demink16, marchant16, map17, chruslinska18, kruckow18, shao18, spera2019} or also \citealt{map18} for a recent review). 
Furthermore, theoretical insights on the most likely properties of the host galaxies provide us with astrophysically-motivated criteria for host localization. These criteria can facilitate the low-latency search for host candidates, by identifying the most probable hosts among the galaxies inside the error box \citep{delpozzo18}. In addition, they can be used to ``rank'' host candidates for those GW events in which we do not observe any electromagnetic counterpart \citep{mapgiactoff18}.

Recent studies \citep{map17,belczynski16,lamberts16,schneider17,marassi19} focused on reconstructing the environment of GW150914 (see \citealt{abbott16d}) by accounting for the evolution of star formation rate density and metallicity ($Z$) across cosmic time. They all agreed that GW150914 formed from metal-poor progenitors ($Z\lesssim 0.1 \ \rmzs$) either at low ($z\sim0.2$) or high redshift ($z\gtrsim 2$). Not only GW150914, but also the other merging binaries which are detected by GW detectors must not necessarily have  formed in recent epochs: they may have  formed in the early Universe and have merged after a long delay time \citep{oshaughnessy10,dominik13,dominik15,map19}. Hence, it is extremely important to learn when and in which circumstances these binaries did form. 


Focusing on the host galaxies of BHBs, \cite{schneider17} suggest that most progenitors of GW150914-like events formed in dwarf galaxies (with stellar mass $M \lesssim 5\times 10^6 \ \rmms$) while their merger occurred while they were hosted in a more massive galaxy ($M>10^{10} \ \rmms$). 
In contrast, DNSs  are expected to form more likely in massive galaxies ($\gtrsim 10^{9} \ \rmms$) at low redshift ($z\lesssim 0.024$)  and tend to merge in the same galaxy where they formed \citep{oshaughnessy10,mapgiactoff18,artale19}. The explanation of this trend is that DNSs are poorly sensitive to progenitor stars' metallicity, while BHBs form preferentially from metal-poor stars  \citep{dominik13, giacmap18a, mapgiac18, perna02}.

This paper is a follow-up of \cite{mapgiactoff18}. Following the same numerical approach as \cite{mapgiactoff18}, we interfaced the Illustris-1 cosmological simulation \citep{vogelsberger14} with  catalogues of merging BHBs, DNSs and neutron star - black hole  binaries (NSBHs) simulated with the {\sc mobse} population-synthesis code \citep{giacobbo18}. While \cite{mapgiactoff18} focused only on compact objects merging in the local Universe ($z\leq{}0.024$), here we extend their analysis to compact objects merging across cosmic time. In particular, we explore  the characteristics of the host galaxies of compact binaries at redshifts $z\sim{}0$, $2$, $4$ and $6$. In addition, we explore how these properties change across the cosmic time and  search for   possible relevant correlations  that could arise among them. This information  will be crucial in the next years with the upgrade of the advanced LIGO/Virgo interferometers to design sensitivity 
and for the next-generation ground-based GW detectors (Einstein Telescope, Cosmic Explorer; \citealt{sathyaprakash12,dwyer15}).



\section{Methods}\label{sec:methods}
\subsection{\sc mobse}
{\sc mobse} \citep{giacobbo18} is an upgrade of the  population-synthesis code {\sc bse} \citep{hurley02}, which includes recent models of mass loss by stellar winds \citep{vink01, vink05}, pair-instability and pulsational pair-instability supernovae (SNe; \citealt{spera17, woosley17}), core-collapse SNe \citep{fryer12}, electron-capture SNe \citep{giacmap18b}. 
Specifically, mass loss is described by an exponential law $\dot{M}\propto Z^{\beta}$, where $\beta=0.85$ for Eddington ratio $\Gamma\leqslant 2/3$, $\beta=2.45-2.4\Gamma$ for $2/3\leqslant\Gamma<1$, and $\beta=0.05$ for $\Gamma>1$ \citep{chen15}. $\Gamma$ is the electron-scattering Eddington ratio, defined as \citep{grafener11}
\begin{equation}
  \log\Gamma=-4.813 + \log(1+X_H)+\log(L_{\star}/L_{\odot})-\log(M_{\star}/M_{\odot}),
\end{equation}
where $X_H$ is the Hydrogen fraction and $L_{\star}$ and $M_{\star}$ are respectively the star luminosity and the star mass. 
The mass distribution of BHs predicted by {\sc mobse} shows a strong dependence on  metallicity: \citet{giacobbo18} obtained BHs with maximum mass of the order of $60 \ \rmms$ for $Z\leqslant 0.001 \ \rmzs$, while the BH maximum mass at solar metallicity ($\rmzs=0.02$) is  $~25 \ \rmms$.
The statistics and in particular the cosmic merger rate of compact-object binaries simulated by {\sc mobse} are in agreement \citep{map17} with the values inferred by the LVC \citep{abbott16a, abbott17c,abbott18a,abbott18b}. For a more detailed description of {\sc mobse} see \cite{giacobbo18} and \cite{giacmap18b}.

In this paper we use  the catalogue of merging compact objects from run $\rm{CC15\alpha5}$ of \cite{giacmap18b}.  
This simulation, in which a high value ($\alpha=5$) for the efficiency of  common-envelope ejection and a low value for SN kicks ($\sigma=15$\,km\,s$^{-1}$) are assumed, is the one which best matches the cosmic merger rate density of DNSs inferred by the LVC \citep{mapgiac18}. Moreover, it matches the expected merger rate of DNSs in the Milky Way, estimated by \cite{pol19}. Run $\rm{CC15\alpha5}$ consists of $1.2\times 10^8$ stellar binaries, divided in 12 sub-sets, corresponding to  different metallicities ($Z= 0.01$, 0.02, 0.04, 0.06, 0.08, 0.1, 0.2, 0.3, 0.4, 0.6, 0.8, and 1.0 $\rmzs$ -- here we assume $Z_\odot{}=0.02$). We expect that our main results are not dramatically affected by the choice of run $\rm{CC15\alpha5}$, since the other runs presented in \cite{giacmap18b} show a very similar trend with metallicity.

\subsection{The Illustris}
We convolve the outputs of {\sc mobse} with the Illustris-1 simulation (hereafter, Illustris), which is the highest resolution run of the Illustris hydrodynamical cosmological simulation project \citep{vogelsberger14,vogelsberger14b,nelson15}. The simulation box's length is 106.5 Mpc, for an overall comoving volume of $(106.5)^3 \ \rm{Mpc^3}$, with an initial baryonic and dark matter mass resolution of $1.26\times 10^6$ and $6.26\times 10^6$~M$_\odot$, respectively. The Illustris box's size provides good resolution for massive haloes, while dwarf galaxies are predominantly unresolved. 
For more detailed  technical  information on the Illustris see \cite{vogelsberger14}.

The sub-grid physical models adopted in the Illustris produce a mass-metallicity relation  \citep{genel14, genel16, vogelsberger14} which is steeper than the relation obtained through observational data \citep{torrey14} and which  does not show the turnover at stellar masses $\gtrsim 10^{10} \ \rmms$. 
In this paper, we thus substitute the simulation's intrinsic mass-metallicity relation with the one expected by observational data. In particular, we use the formula by \cite{maiolino08} and \cite{mannucci09}:
 
 \begin{equation}\label{eq:maiolino}
 12+\log{({\rm O}/{\rm H})}=-0.0864\,{}(\log{M} - \log{M_0})^2\,{}+K_0,
 \end{equation} 
where $M$ is total stellar mass of the host galaxy in solar mass units, $M_0$ and $K_0$ are parameters determined at each redshift by best-fittings with observed data points (Table \ref{tab:maiolino}).

\begin{table}
	\caption{Best-fitting parameters for the mass-metallicity relation  obtained by Maiolino et al. (2008) at different redshifts.}
	\begin{center}
		
				\begin{tabular}{c c c}
					
					\hline
					$z$ & $\log M_0$ & $K_0$ \\
					\hline
					0.07 & 11.18 & 9.04 \\
					0.7 & 11.57 & 9.04\\
					2.2 & 12.38 & 8.99\\
					3.5 & 12.28 & 8.69 \\
					\hline
			\end{tabular}

	\end{center}

	\label{tab:maiolino}
\end{table}

In equation~\ref{eq:maiolino}, for redshifts $z>3.5$ or $z<0.07$ we use the same coefficients as for   $z=3.5$ and $z=0.07$, respectively; for redshifts between the ones listed in the table we adopt a linear interpolation. We extract the metallicity of each Illustris particle from a Gaussian distribution with mean value given by equation \ref{eq:maiolino} and standard deviation $\sigma=0.3$  dex to account for metallicity dispersion within galaxies. By repeating the calculations for a larger ($\sigma=0.5$) and a smaller ($\sigma=0.2$) scatter,  we verified that our results are unchanged  (see also \citealt{map17,mapgiactoff18}).  

\subsection{The Monte Carlo algorithm}
We combined the catalogue of simulated compact-object binaries by {\sc mobse} with the Illustris cosmological simulation through a Monte Carlo algorithm as follows. First, we extract the initial mass $M_{\rm Ill}$, metallicity $Z_{\rm Ill}$ and formation redshift $z_{\rm Ill}$ of each stellar particle in the Illustris. Second, we catalogue the masses of  compact objects  simulated by {\sc mobse} and their delay time $t_{\rm delay}$ (i.e. the  time elapsed between the formation of their progenitor stars and the merger). 
Finally, we use the following algorithm 

\begin{equation}\label{eq:planting}
N_{\rm CO, i}=N_{\rm BSE,i}(Z\sim{}Z_{\rm Ill})\,{}\frac{M_{\rm Ill}(Z_{\rm Ill})}{M_{\rm BSE}(Z\sim{}Z_{\rm Ill})}\,{}f_{\rm corr}\,{}f_{\rm bin}
\end{equation}

to associate a number $N_{\rm CO, i}$ of simulated compact-object binaries to each  Illustris stellar particle (where $i$ = BHB, NSBH or DNS indicates the type of compact object). In equation~\ref{eq:planting}, $M_{\rm \rm BSE}(Z\sim{}Z_{\rm Ill})$ is the total initial mass  of the stellar population simulated with {\sc mobse} which has metallicity $Z$ closer to the one of the Illustris particle $Z_{\rm Ill}$ (with $Z$ chosen among the 12 metallicities simulated with {\sc mobse}), $N_{\rm BSE, i}(Z\sim{}Z_{\rm Ill})$ is the number of merging compact objects associated with the population of mass $M_{\rm \rm BSE}(Z\sim{}Z_{\rm Ill})$, $f_{\rm bin}=0.5$ is the assumed binary fraction and $f_{\rm corr}=0.285$ is a correction factor (to take into account that run~CC15$\alpha{}5$ contains only primary stars with zero-age main-sequence mass $M_{\rm ZAMS}>5 \ \rmms$). 

We estimate for each compact object binary the lookback time at which it merges as  $t_{\rm merg}=t_{\rm form}-t_{\rm delay}$, where $t_{\rm form}$ is the lookback time at which the host Illustris particle formed:
\begin{equation}\label{eq:tform}
t_{\rm form}=\frac{1}{H_0}\int_{0}^{z_{\rm Ill}}\frac{1}{(1+z)\,{}[\Omega_{\rm M}\,{}(1+z)^3+\Omega_\Lambda]^{1/2}}dz,
\end{equation}
where the cosmological parameters $H_0=100h$\,km\,s$^{-1}$ ($h=0.704$), $\Omega_\Lambda=0.7274$ and $\Omega_{\rm M}=0.2726$ are the ones adopted in the Illustris \citep{hinshaw13}.


\section{Results}
Through the formalism described in the previous section, we can analyze the main properties of the environment of merging compact objects, which are the stellar  mass of host galaxy where the binary has formed ($M_{\rm form}$), the stellar mass of the host galaxy where the merger occurs ($M_{\rm merg}$), the delay times ($t_{\rm delay}$) and the metallicity $Z$ of the progenitor stars. We consider compact objects which reach coalescence in the Illustris snapshots corresponding to the redshift intervals $0.01<z<0.02$ (hereafter, $z\sim{}0$), $2.00<z<2.10$ (hereafter, $z\sim{}2$), $4.01<z<4.43$ (hereafter, $z\sim{}4$) and $6.01<z<6.14$ (hereafter, $z\sim{}6$) and formed previously. 



\begin{figure*}
	\begin{center}
		\includegraphics[scale=0.47]{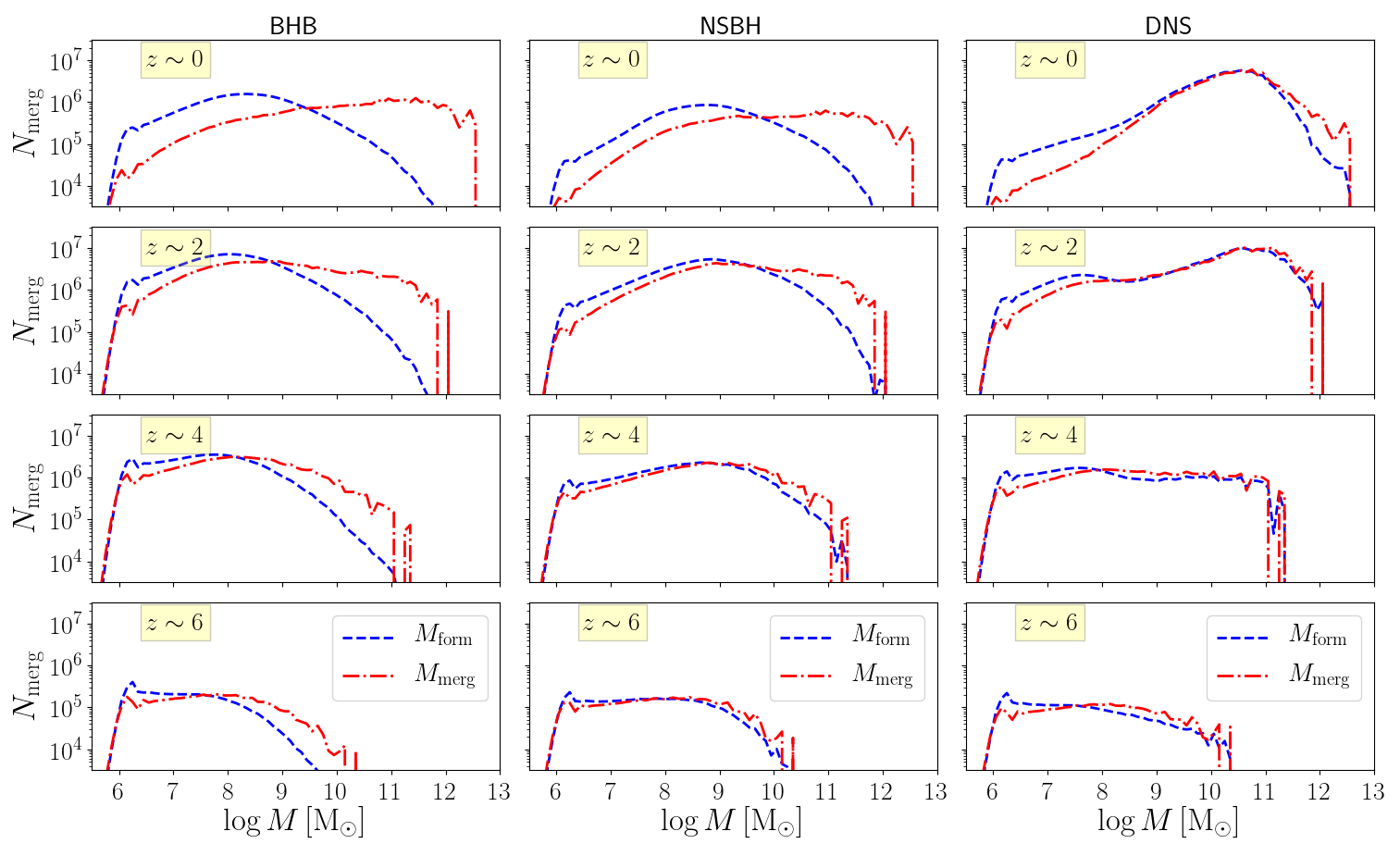}
	\end{center}

        \caption{Distribution of the stellar mass of the host galaxies where the compact binaries merge ($\mm$, red dot-dashed lines) and of the stellar mass of the host galaxies where where their stellar progenitors formed ($\mf$, blue dashed line). Left-hand column: BHBs; middle column: NSBHs; right-hand column: DNSs. From top to bottom: compact binaries merging at redshift $z\leq{}0.01$, $z=2.0$, $z=4.0$ and $z=6.0$, respectively.}
\label{fig:histmasses}
\end{figure*}

Figure \ref{fig:histmasses} shows the distribution of $\mf$ and $\mm$ for those BHBs, NSBHs and DNSs which have formed at $z>0.01$, $z>2$, $z>4$, $z>6$ and have merged in the Illustris snapshot corresponding to $z\sim{}0$, $z\sim{}2$, $z\sim{}4$ and $z\sim{}6$.  At low redshifts, the mass $\mf$ of the formation host of BHBs and NSBHs  is  typically lower than the mass $\mm$ of the galaxy where the merger occurred. 
This is not true for DNSs: both $\mf$ and $\mm$ distributions peak around the same range (about $10^9 -10^{11} \ \rmms$) even in the $z\sim{}0$ case. These trends were already discussed in \cite{mapgiactoff18}: most BHBs and NSBHs merging at low $z$ arise from metal-poor progenitors at high $z$. Thus, BHBs and NSBHs tend to form in smaller galaxies and to merge in larger ones, because of the hierarchical assembly of galaxies. In contrast, DNSs merging in the local Universe form predominantly from metal-rich progenitors with short delay time. Thus, they tend to form and merge in the same galaxy.  

At $z\sim{}2$, the difference between the  distribution of $\mf$ and $\mm$ becomes less marked and the maximum  host mass is lower. This happens for two reasons: firstly, the time elapsed from the Big Bang 
to $z=2$ is $\sim{}3$ Gyr, so the largest galaxies did not have enough time to form (according to the hierarchical clustering model of galaxy assembly); secondly, following the same line of thought, even the delay time from the formation of a binary to its merger must be shorter 
than the $z\sim 0$ case and thus the mass of the formation host is closer to the mass of the merger host.

At $z\sim 4$, this trend is even stronger. In particular, almost all DNSs and most NSBHs form and merge in the same host galaxy.  The difference between the distribution of $\mf$ and that of $\mm$ is significantly smaller  even in the case of BHBs.
Moreover, the distribution of DNS host masses, which peaked around $\mf\sim{}\mm{}\sim{}10^{11}$ M$_\odot$ at $z\sim{}0-2$, is now rather flat. 

The number of compact objects merging at $z\sim{}6$ is significantly lower than at lower redshifts. While the $\mf$ and $\mm$ distributions for NSBHs and DNSs are now almost identical,  for BHBs
we still see a small difference between $\mf$ and $\mm$. 

Galaxies with stellar mass $\lesssim 10^6 \ \rmms$ are associated with just one Illustris stellar particle: smaller galaxies cannot form in the Illustris (see \citealt{schneider17} for a different numerical approach which enables to study smaller dwarf galaxies). 
The spike we see in the $z\sim2$ ($z\sim4$) case around $\log{M/\rmms}=12$ ($\log {M/\rmms}=11.5$) is explained as a single massive galaxy being formed in the Illustris simulation.
Finally, we notice that at $z\sim2$ we have a higher total number of mergers compared to the other cases, due to the peak of star formation in this redshift range \citep{madau14}. This trend is in agreement with previous papers \citep{dominik13,dominik15,map17,mapgiac18,eldridge19}. 

%
%
%
%
%
%
%
%
%
%
%
%
\begin{figure*}
	
	\subfigure{\includegraphics[width=0.495\textwidth]{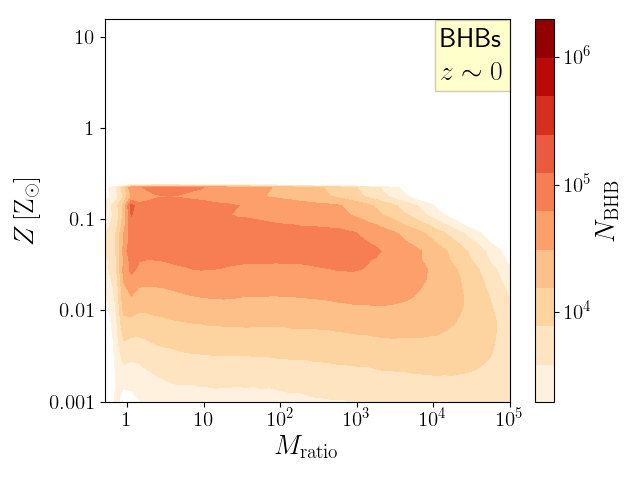}}
	\subfigure{\includegraphics[width=0.495\textwidth]{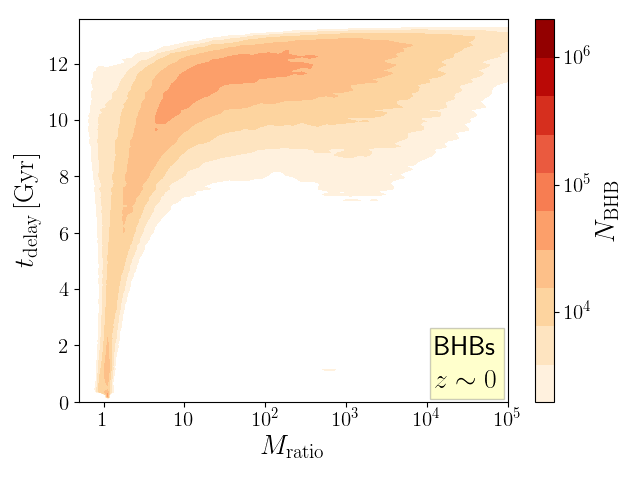}}
	\subfigure{\includegraphics[width=0.495\textwidth]{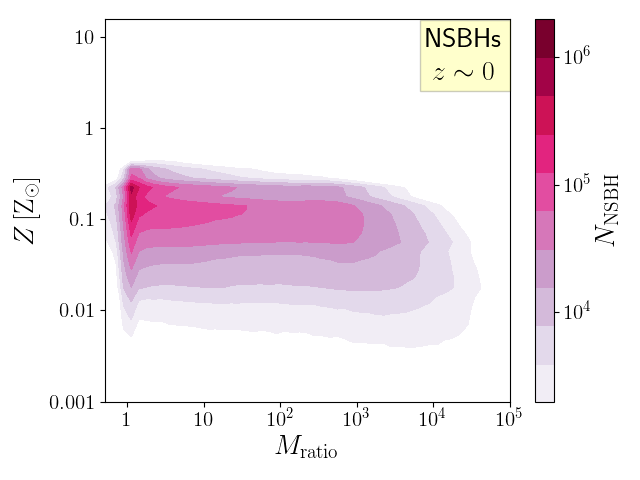}}
	\subfigure{\includegraphics[width=0.495\textwidth]{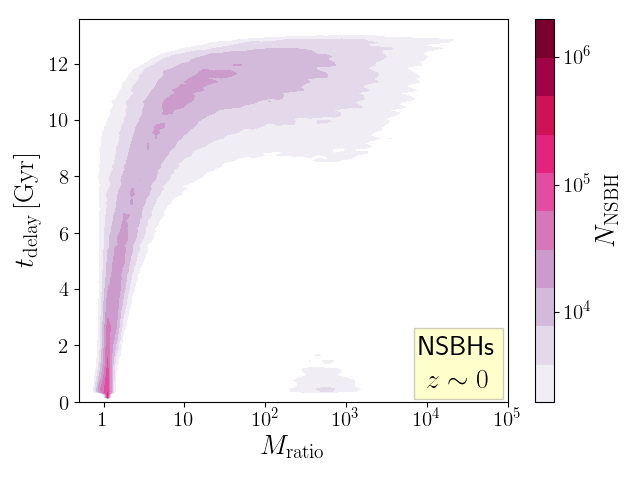}}
	\subfigure{\includegraphics[width=0.495\textwidth]{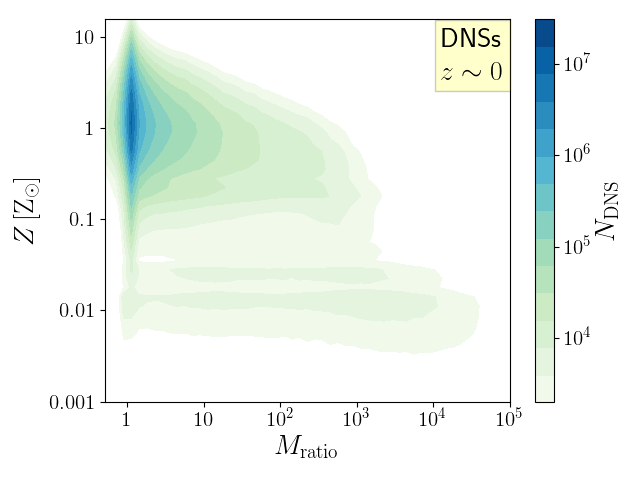}}
	\subfigure{\includegraphics[width=0.495\textwidth]{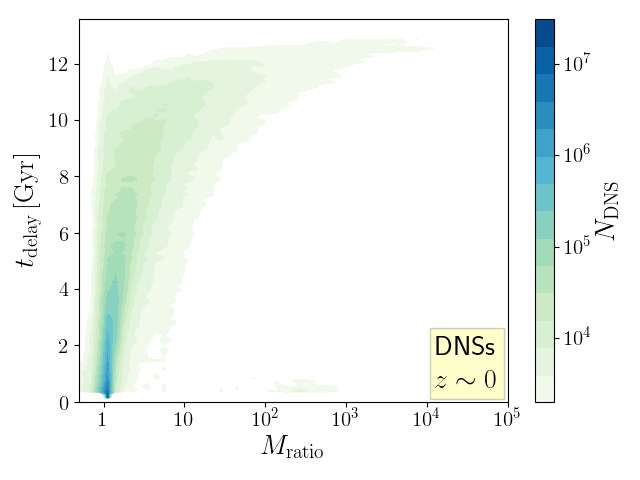}}
		
	\caption{Left-hand panel: progenitor's metallicity versus ratio $\mr=M_{\rm merg}/M_{\rm form}$ for DNSs (blue), NSBH (purple) and BHBs (orange) merging at $z\leq{}0.01$; right-hand panel:  delay time $t_{\rm delay}$ versus $\mr$ for DNSs (blue), NSBH (purple) and BHBs (orange) merging at $z\leq{}0.01$. The colour-coded map (in logarithmic scale) indicates the number of merging compact objects per cell. Cells with $<2\times{}10^3$ compact objects binaries are shown in white.
		 The cell sizes are $\log \delta Z/Z_{\odot} \times \log \delta \mr/M_{\odot} = 0.1\times 0.1$ and $\log \delta Z/Z_{\odot} \times \log \delta \td/\td=0.1\times 0.1$ for $\mr - Z$ and $\td - Z$ plots, respectively.  Contour levels are spaced by a number of $10^{0.3}$ binaries each other.  }
	\label{fig:bhbdns134}
	
\end{figure*}

We now look for possible correlations between host galaxy mass,  delay time and  metallicity of the compact binary. The left-hand panels of Figure~\ref{fig:bhbdns134} show the metallicity $Z$ of the progenitor as a function of the mass ratio $\mr = M_{\rm merg}/M_{\rm form}$, while the right-hand panels show the delay time $\td$ between the formation of the binary and the merger  as a function of the mass ratio $\mr$,  for BHBs, NSBHs and DNSs merging at $z=0$.

It is apparent that DNSs lie in a different region of the $\mr-Z$ plot with respect to BHBs and NSBHs.
At $z\sim{}0$, the vast majority of DNSs are characterized by a host mass ratio $\sim 1$, suggesting that they formed and merged in the same galaxy, and by a progenitor metallicity in the range $\sim 0.1 - 10 \ \rmzs$. In contrast, the host mass ratio for BHBs spans mainly over a large interval $\mr \sim 1 - 10^{4}$  (i.e. $M_{\rm merg}\ge{}M_{\rm form}$),  and the metallicity of their progenitors is always sub-solar, with a preference for $Z\sim 0.02-0.2 \ \rmzs$. NSBHs occupy approximately the same region as BHBs. As the latter, they never show a super-solar metallicity; on the other hand, a  stronger preference for a smaller range of $\mr$ (mainly $\sim 1 - 10^{3}$) is visible compared to BHBs. The number of merging NSBHs characterized by a $Z\le 0.01 \ \rmzs$ is very low (less than $<2\times{}10^3$).

We interpret this difference in terms of delay times (right-hand panels of Figure~\ref{fig:bhbdns134}): most BHBs merging at $z\sim{}0$ form in smaller galaxies at high redshift and then merge at low redshift with a long delay time. Given the long BHB $t_{\rm delay}$, the initial host galaxy  had enough time to grow in mass because of galaxy mergers and accretion. In contrast, most DNSs merging at $z\sim{}0$ form in nearby galaxies and merge in the same galaxy with a short delay time. At $z\sim{}0$ NSBHs mainly form in small galaxies at high redshift and merge after a long delay time, but we find also a good number of them forming in closer galaxies and merging within a shorter delay time.
This interpretation is supported by the right-hand panel of Figure~\ref{fig:bhbdns134}, showing $t_{\rm delay}$  as a function of $\mr$. The majority ($\sim{}80~\%$) of DNSs merge within a $t_{\rm delay}<4$ Gyr, while 65~\% of BHBs have $t_{\rm delay}>10$ Gyr.
Around 20~\% of NSBHs form and reach coalescence  within $\td \leq 4$ Gyr, while the $\sim{}40$~\% of them have $t_{\rm delay}>10$ Gyr.

\begin{figure*}
	\subfigure{\includegraphics[width=0.495\textwidth]{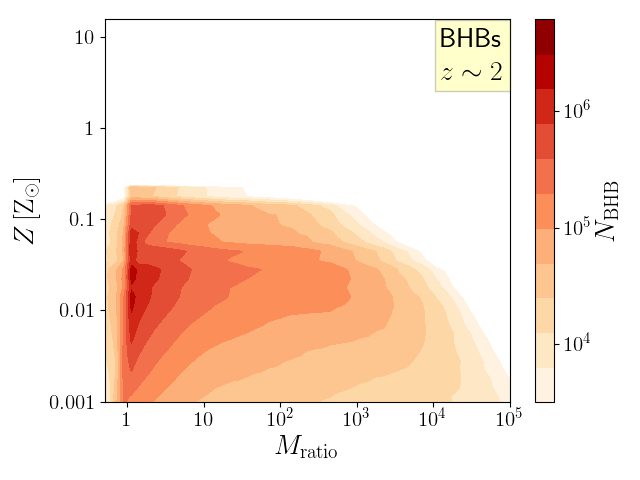}}
	\subfigure{\includegraphics[width=0.495\textwidth]{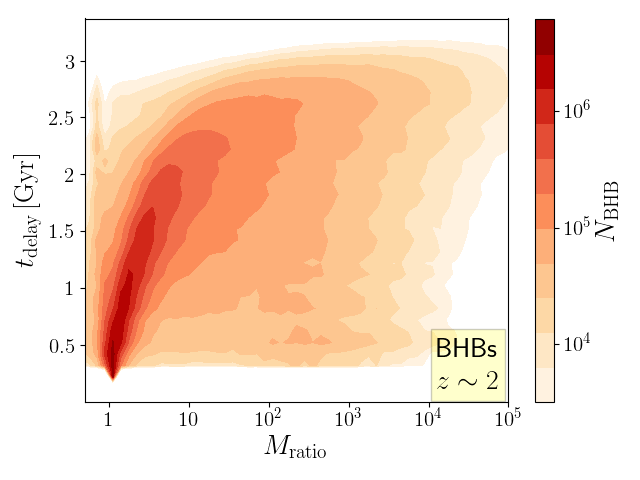}}
	\subfigure{\includegraphics[width=0.495\textwidth]{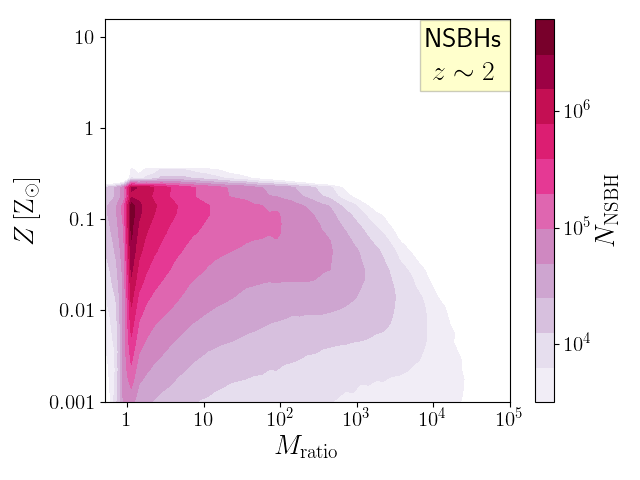}}
	\subfigure{\includegraphics[width=0.495\textwidth]{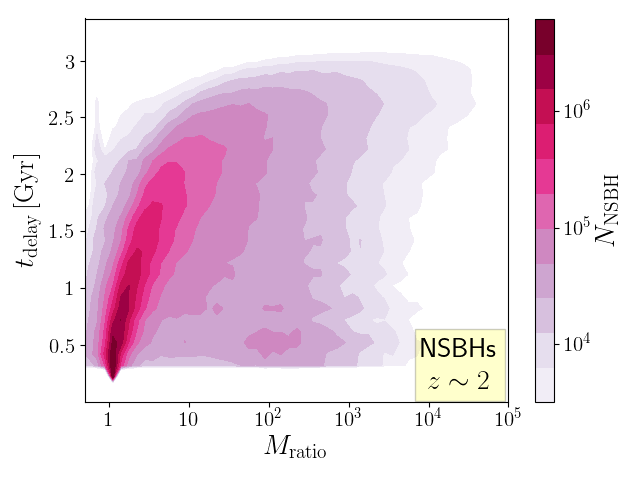}}
	\subfigure{\includegraphics[width=0.495\textwidth]{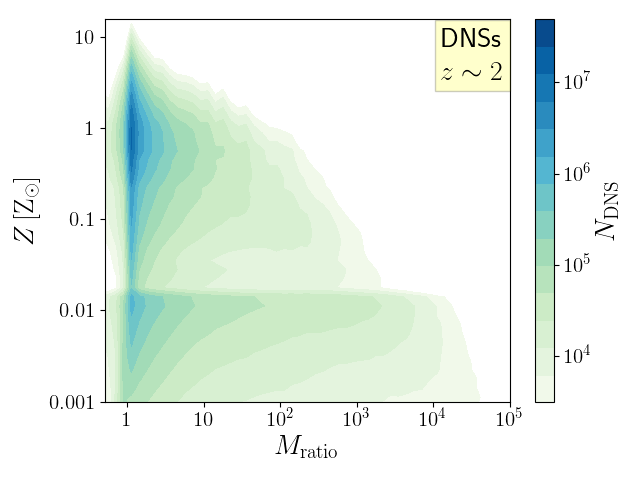}}
	\subfigure{\includegraphics[width=0.495\textwidth]{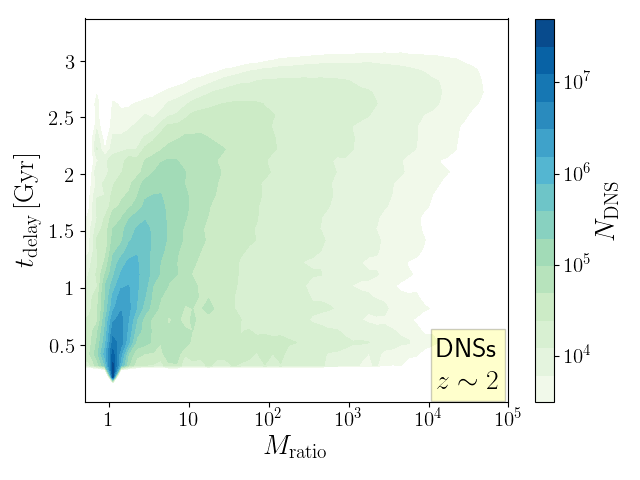}}

	\caption{Same as Figure~\ref{fig:bhbdns134}, but for DNSs (bottom), NSBH (middle) and BHBs (top) merging at redshift $z\sim{}2$.
	Cells with $<3\times10^3$ compact objects binaries are shown in white.}
	
	\label{fig:bhbdns68}
	
\end{figure*}

Figure \ref{fig:bhbdns68} is the same as Figure~\ref{fig:bhbdns134} but for compact-object binaries merging at $z\sim{}2$ (lookback time: 10.39 Gyr). 
  In the upper panels, we notice that BHBs which merge at $z\sim 2$ are mainly characterized by a $\mr$ of the order of tens. More specifically, $\sim{}50$~\% of  them merge within a mass ratio interval of $1-24$, as reported in Table \ref{tab:mratio}.  
The metallicity of BHB progenitors is always sub-solar and in particular we have more progenitors with $Z<0.01 \ \rmzs$ with respect to the $z\sim 0 $ case. 

DNSs merging at $z\sim{}2$ still show a preference for $\mr\sim{}1$, but a significant number of objects have $\mr\gg{}1$. In particular, DNSs which form from metal-poor progenitors tend to have larger values of $\mr$. While most  DNSs merging at $z\sim{}2$ have solar metallicity progenitors, the fraction of metal-poor progenitors increases significantly with respect to DNSs merging at $z\sim{}0$. Metal poor progenitors are also associated with slightly longer delay times.

From Table \ref{tab:mratio} we see that about half of NSBHs merging at $z\sim{}2$ are characterized by a $\mr$ of the order of one and are thus associated with a short delay time. The remaining half are spread in a $\mr$ interval spanning from $\sim{}10$ to $\sim{}10^4$.
Likewise DNSs, the number of metal-poor progenitors sensibly increased with respect to the $z\sim{}0$ case.

\begin{figure*}
	
	\subfigure{\includegraphics[width=0.495\textwidth]{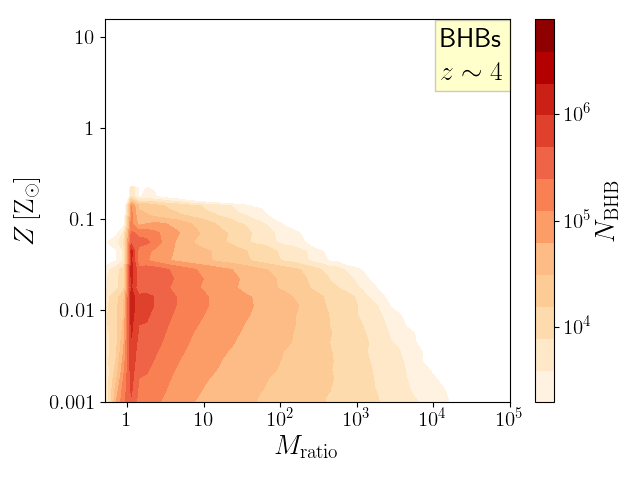}}
	\subfigure{\includegraphics[width=0.495\textwidth]{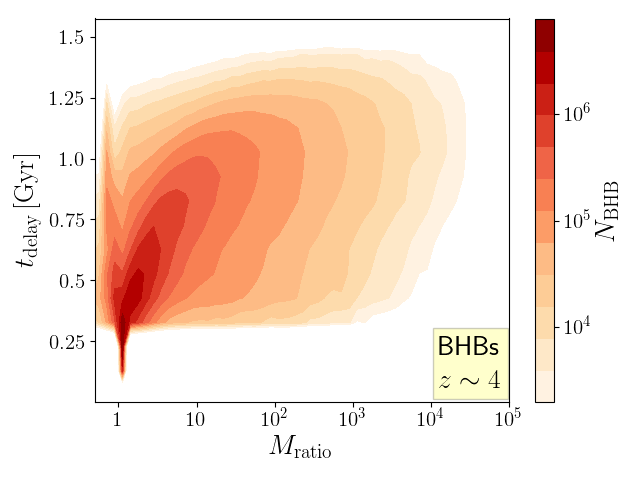}}
	\subfigure{\includegraphics[width=0.495\textwidth]{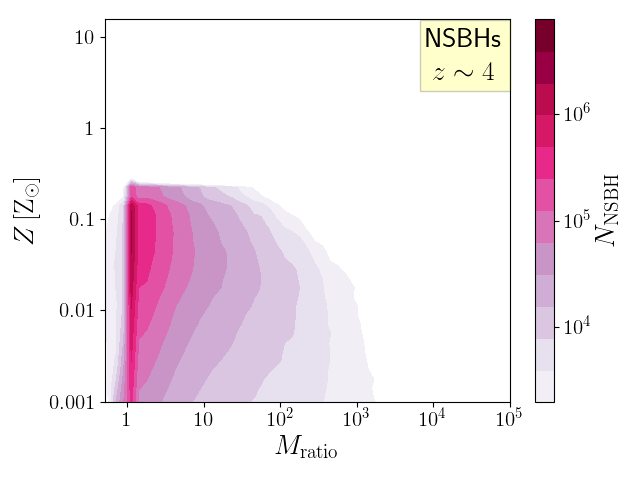}}
	\subfigure{\includegraphics[width=0.495\textwidth]{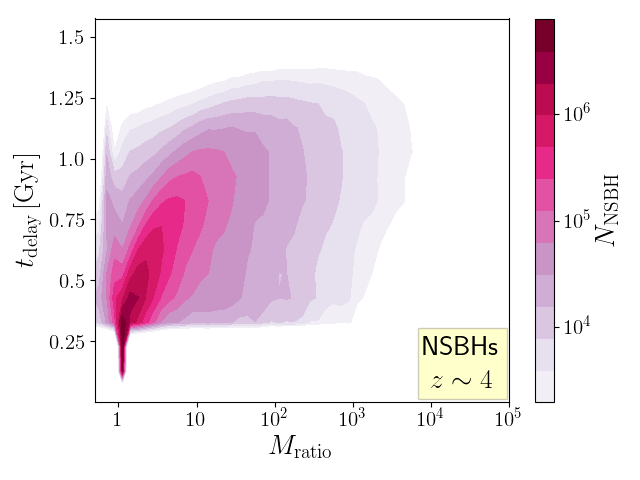}}
	\subfigure{\includegraphics[width=0.495\textwidth]{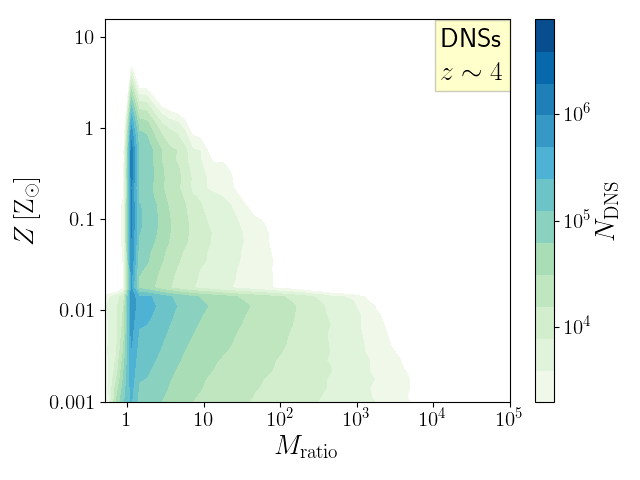}}
	\subfigure{\includegraphics[width=0.495\textwidth]{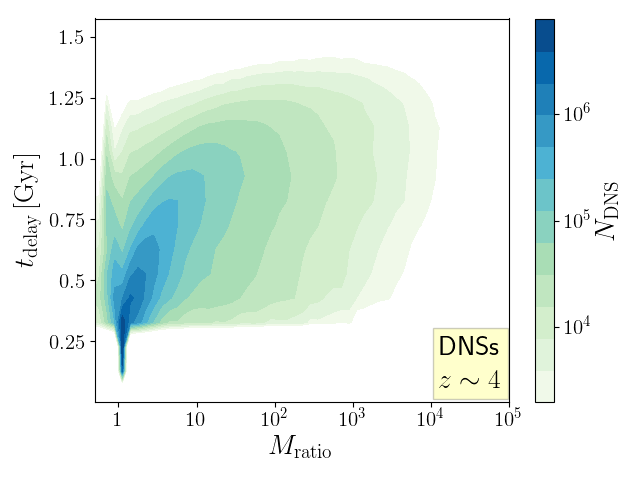}}

	\caption{Same as Figure~\ref{fig:bhbdns134}, but for DNSs (bottom), NSBH (middle) and BHBs (top) merging at redshift $z\sim{}4$.} 
	\label{fig:bhbdns54}
	
\end{figure*}

Figure~\ref{fig:bhbdns54} is the same as Figure~\ref{fig:bhbdns134} but for compact object binaries merging at $z\sim4$ (lookback time: 12.17 Gyr). 
BHB progenitors squeeze to even lower metallicity. In particular, the vast majority of systems have $0.001\le{}Z/\rmzs{}\le{}0.1$. Host mass ratios are sensibly smaller ($\mr\lesssim{}6$ in the  50\% of cases), as it is reasonable to expect because the Universe is $\lesssim{}1.5$ Gyr old.

The progenitors of DNSs merging at $z\sim{}4$ shift to  lower metallicities with respect to those merging at $z\sim{}2$. The DNSs with metal-poor progenitors tend to have larger values of $\mr$ and longer delay times than the DNSs with metal-rich progenitors. Hence, DNSs with metal-poor progenitors seem to behave more like BHBs than like the other DNSs. 

Table~\ref{tab:mratio} shows that 50\% of NSBHs merging at $z\sim{}4$ have a narrow host mass ratio range, around $\sim{}1-2.4$, which is related to a very short delay time.

\begin{figure*}
	\subfigure{\includegraphics[width=0.495\textwidth]{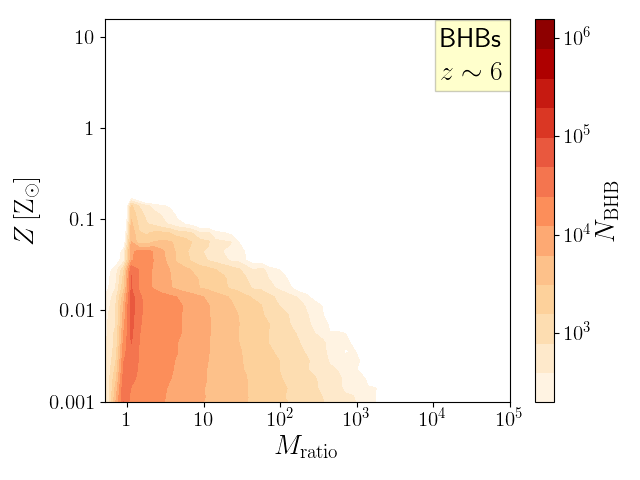}}
	\subfigure{\includegraphics[width=0.495\textwidth]{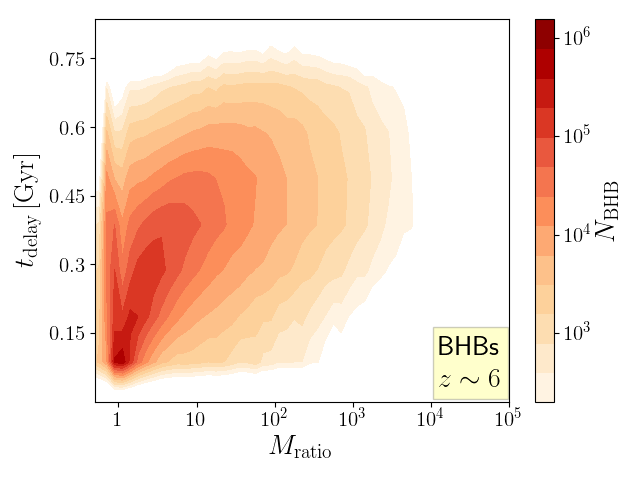}}
	\subfigure{\includegraphics[width=0.495\textwidth]{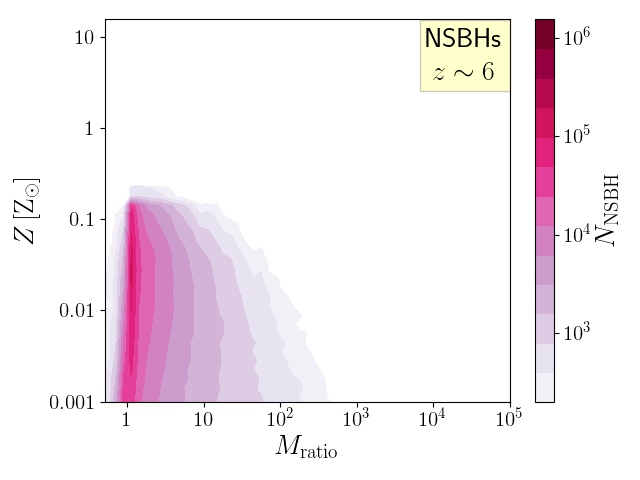}}
	\subfigure{\includegraphics[width=0.495\textwidth]{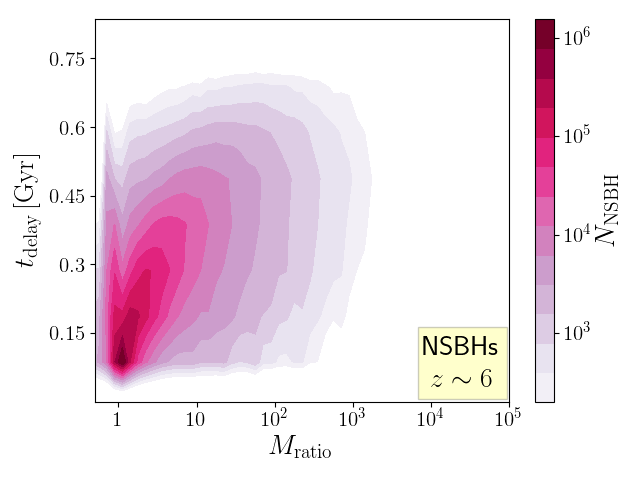}}
	\subfigure{\includegraphics[width=0.495\textwidth]{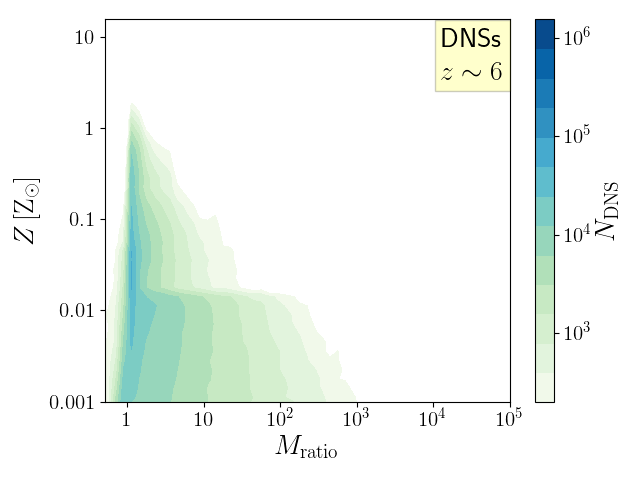}}
	\subfigure{\includegraphics[width=0.495\textwidth]{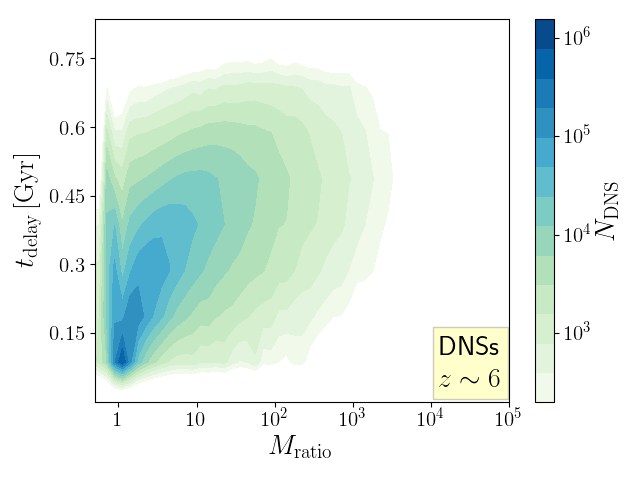}}

	\caption{Same as Figure~\ref{fig:bhbdns134} but for DNSs (bottom), NSBH (middle) and BHBs (top) merging at $z\sim{}6$. Cells with $<3\times10^2$ compact objects binaries are shown in white.}
	\label{fig:bhbdns45}
	
\end{figure*}

Finally, Figure \ref{fig:bhbdns45} shows BHBs, NSBHs and DNSs merging at $z\sim6$ (lookback time: 12.797 Gyr). At such high redshift the number of mergers is much smaller than in the previous cases, because the cosmic star formation rate decreases quite fast for $z\ge{}4$. The most noticeable feature in this Figure is that all compact object binaries merging at $z\sim{}6$ are very similar populations in terms of $\mr$ and $t_{\rm delay}$.

To summarize, our main result is that BHBs merging at low redshift are a significantly different population from DNSs merging at low redshift: the former have larger values of $\mr$, much smaller progenitor's metallicity and much longer $t_{\rm delay}$ than the latter. As redshift increases, this difference becomes smaller: a sub-population of DNSs with metal-poor progenitors and longer delay times
appears at $z\sim{}2-4$, well distinguished from the bulk of DNSs with metal-rich progenitors and short delay times. At high redshift $z\sim{}6$, the population of merging DNSs is quite similar to the population of merging BHBs: both of them have mostly metal-poor progenitors and short delay times. 
The population of NSBHs shows a similar trend to BHBs at low redshift,  in terms of $\mr$, $\td$ and progenitor's metallicity. With increasing redshift,  $\mr$ and $\td$ shrink to values similar to DNSs, rather than BHBs. At $z\geq{}2$ the main difference between NSBHs and DNSs is  the metallicity range of progenitors.

\begin{table}
\caption{Mass ratio interval (in $\rmms$) within which the 50\% of compact objects binaries merge at the distinct investigated redshifts.}
\begin{center}
	
	\begin{tabular}{c c c c c}
		
		\hline
		 			& $z\sim0$ & $z\sim2$ & $z\sim4$ & $z\sim6$ \\
		\hline
		$M_{\mathrm{ratio, BHB}}$ & $6.1-7.2\times{}10^2$	&$1.3-24$	&$1-6.1$	&$1-5$	\\
		$M_{\mathrm{ratio, NSBH}}$& $1.5-1.6\times{}10^2$ 	&$1-5.8$	&$1-2.4$	&$0.9-1.9$	\\
		$M_{\mathrm{ratio, DNS}}$ & $1-2.7$ 	&$0.9-1.5$	&$1-2.7$	&	$1-3.5$\\
	
		\hline
	\end{tabular}

\end{center}

\label{tab:mratio}
\end{table}

%
%
%
%
%
%
%

\section{Discussion and Conclusions}

In this work, we have characterized the environment of merging compact objects across cosmic time, by means of population-synthesis simulations (run with the code {\sc mobse}, \citealt{giacmap18a,giacobbo18}), combined with the Illustris cosmological simulation \citep{vogelsberger14,vogelsberger14b,nelson15} through a Monte Carlo algorithm \citep{map17,map18}. 

We focused on the stellar mass of the host galaxy where the compact binaries merge, the stellar mass of the host galaxy where the progenitor stars formed, the metallicity of progenitor stars and the delay time between the formation and the merger. We studied three types of compact objects binaries, black hole-black hole (BHBs), neutron star-black hole (NSBHs) and neutron star-neutron star binaries (DNSs), for four reference epochs, corresponding to redshift $0,2,4$ and $6$.

Present-day merging BHBs mainly formed  $\sim{}10-12$ Gyr ago and merge in a more massive host galaxy than the one where they formed, due to the hierarchical clustering build-up process. In general, BHBs need a longer delay time to reach the merger phase compared to other compact-object binaries, even at higher redshifts. BHBs which merge nowadays have likely formed in galaxies with mass $\sim 10^7 - 10^{10} \ \rmms$ from metal-poor progenitors, but they merge in galaxies of every mass in the range $10^8 -10^{12} \ \rmms$. 

The difference between the mass of the formation host and the mass of the merger host becomes smaller and smaller as redshift increases (see Figure~\ref{fig:histmasses}). Lower masses of the formation hosts and lower metallicities of BHB progenitors are more likely going back in time, because smaller galaxies and metal-poor stars were more common in the past. This is a general trend,  visible also for NSBHs and DNSs.

DNSs  have a different behaviour with respect to BHBs. DNSs  have a strong predisposition for  high progenitor metallicities at any epochs (up to many times the solar value), and are characterized by low values of $\mr$, associated with short delay times.
 A secondary population of DNSs starts to emerge at $z\gtrsim{}2$. This sub-population is characterized by higher values of $\mr$ and $\td$ and metal-poor progenitors, similar to BHBs.

 The current DNS host mass range ($10^9-10^{11} \ \rmms$) is consistent with the mass range of short gamma-ray burst hosts \citep{fong13}, strengthening the link between DNS mergers and SGBRs.

We foresee to find merging NSBHs at the present day mainly in galaxies within a  mass range of $\sim 10^{7.5} - 10^{10.5} \ \rmms$. The population of  NSBHs merging at $z\sim{}0$ shows a preference for small $\mr$, thus a tendency for shorter delay times; nonetheless, we find that many NSBH mergers are characterized by delay times of the order of $10$ Gyr. At higher redshifts, $\mr$ and $\td$ squeeze to lower values, similar to DNSs. NSBHs are  characterized by a sub-solar metallicity of progenitors at each investigated epoch.

The $\mr-Z$ plots of DNSs (Figure \ref{fig:bhbdns134}, \ref{fig:bhbdns68}, \ref{fig:bhbdns54} and \ref{fig:bhbdns45}) clearly shows that there are two distinct sub-populations of DNSs: one with metallicity about solar (or above solar) and $\mr$ strongly peaked around one, the other with metallicity $Z<0.02 \ \rmzs$ and with a broader distribution of $\mr$. 


These two sub-populations arise from the population-synthesis model CC15$\alpha{}5$ we adopted. Figure~14 of \cite{giacmap18a} shows that in model CC15$\alpha{}5$ DNSs born from metal-poor ($Z\lesssim{}0.0004$) and metal-rich progenitors ($Z\gtrsim{}0.01$) have approximately the same merger efficiency (defined as the number of mergers that we expect from a coeval stellar population integrated over the Hubble time), while the merger efficiency drops by a factor of $\sim{}5-10$ for  DNSs with intermediate-metallicity progenitors ($Z\approx{}0.002$).

In conclusion, our study suggests that the mass of the host galaxies encodes important information on the populations of merging compact objects across cosmic time. Upcoming third-generation ground-based GW detectors will be able to observe DNS mergers up to $z\ge{}2$ and BHB mergers up to $z\ge{}10$, unveiling the properties of merging compact objects as a function of redshift \citep{kalogera19}. In preparation for the era of third-generation GW detectors, we show that characterizing the environment and the host mass  of compact-object binaries at high redshift is crucial to identify the population of their progenitor stars. 


\section*{Acknowledgements}
 MM  acknowledges financial support by the European Research Council for the ERC Consolidator grant DEMOBLACK, under contract no. 770017. MCA and MM acknowledge financial support from the Austrian National Science Foundation through FWF stand-alone grant P31154-N27 ``Unraveling merging neutron stars and black hole-neutron star binaries with population synthesis simulations''.  MT thanks Michele Ronchi for his support in computational tasks. We thank the internal Virgo referee Barbara Patricelli for her useful comments.




\bibliographystyle{mnras}
\bibliography{bibliography} 







	
	



\bsp	
\label{lastpage}
\end{document}